# A shock wave instability induced on a periodically disturbed interface

A. Markhotok

*Physics Department*
*Old Dominion University*
*Norfolk, VA 23529 USA*

**Abstract**

The shock wave instability induced when interacting with a small waviness on an interface was investigated analytically and numerically. The perturbation to the shock was phenomenologically treated assuming this as the consequence of the shock refraction. The instability develops in the form of wave-like stretchings into the lower density medium followed with the loss of stability in the flow behind it, and eventually evolving into an intense vortex structure. The instability mode is aperiodical and unconditional, and either a transition to another stable state or continuous development as a secondary flow is possible. Among other interesting features are: a similarity law in the spatial and temporal evolution of the perturbations with respect to the interface curvature; the instability locus independence of the gas density distribution thus identifying the interface conditions as the sole triggering factor; the role of the density gradient in the instability evolution discriminating between qualitatively different outcomes; and the possibility of decay via non-viscous dumping mechanisms. The phenomenological connection between the shock and the interface stability is discussed.

*PACS numbers*: 47.32. C- , 47.32. –y, 47.32. Ef, 52.35.We. 47.15.Fe, 47.32.C, 47.20.-k, 47.40.-x, 47.40.Nm

E-mail of corresponding author: amarhotk@phys.washington.edu

## I. INTRODUCTION

The shock wave stability is an important topic of compressible fluid dynamics having applications in a wide range of problems including combustion, shock–flame interactions, energy deposition for high speeds flow control, astrophysics plasmas, and inertial confinement fusion research. Understanding the phenomenology of the instability onset and development is particularly important in interpreting experimental results of the shock-heated medium interactions, where multiple-process phenomena are common and each component in the complex picture needs to be identified. With understanding the key physical mechanisms governing the components, they can be confidently distinguished based on the timing and the scale through estimation of its characteristic parameters.

A weak short-time sinusoidal disturbance applied to an isolated planar shock propagating in a uniform medium that obeys an ideal-gas equation of state is known to decay with time, regardless of whether the viscous damping is present in the flow or not [1,2]. However, if the shock is perturbed via the change of parameters of the media where it propagates, it can amplify resulting in reorganization of its initially stable state in the form of a secondary flow or transition to another stable state. Such a situation can be modeled with a planar shock wave interacting with an interface that was spatially disturbed as a result of, for example, Rayleigh-Taylor (RT) instability or the anisotropic effects having a place during the initial phase of a discharge-induced bubble explosion [3,4].

The topic of the present research will be to investigate the shock stability when it interacts with such a periodically disturbed interface. It will be studied by perturbing an initially planar stable shock via the disturbance of the media parameters in front of it and then looking into conditions necessary to ensure a steady growth of the perturbation. The work will be mainly focused on the phenomenological origin of the shock instability and identifying the set of system parameters critical for its triggering and positive dynamics.

The shock refraction is assumed here as the only mechanism mediating the interaction between the shock and the interface disturbance, so the resulting perturbation to the shock will be phenomenologically treated as the consequence of the processes studied earlier in [5, 6]. In accordance to these and many other studies, for ex. [3,7,8], the shock wave structure and the mode of its propagation may undergo substantial modifications when it crosses a discontinuous interface between two gases with different properties. The transformations can proceed to the extent that the shock's geometrical state changes qualitatively, for example from being planar to almost spherical [5,6], or from spherical to a nearly perfect cone [9,10]. The front perturbation amplitude can be controlled by the mode of the interface disturbance, the initial shock strength, the type and the strength of the gas density distribution [11,12], and the conditions across the interface such as heating intensity and the interface smoothness [13]. Thus it can be said that, as a result of interaction with the interface disturbance, the shock wave loses its stability by irreversibly departing from its initially stable state. The perturbation to the shock will either amplify indefinitely (or until its decay) or eventually evolve to another state of stability characterized by new geometrical and dynamical parameters. Pressure perturbations caused by the front distortions may further result in the down-stream flow (behind the shock) parameter re-distribution that breaks its stability by inducing a rotational motion eventually evolving into an intense donut/roll shaped vortex structure of a considerable size [6].

In the next paragraph, a planar shock wave interacting with the interface that was disturbed previously and became stationary by the time of the shock arrival at its location will be considered. A steady motion of the planar shock wave through a uniform gas with a speed $V_1$ toward the interface will define the initial system's stationary state. The state is stable as all the variables $\{x_1, x_2, \ldots x_n\}$ defining it are not time dependent. The variables will include geometrical parameters (such as form and dimensions of the interface disturbance pattern and the shock front) and dynamical (velocity field, rotation, temperature/density gradients, pressure gradients, etc). The goal is to describe the reaction of the system as the shock starts to interact with the disturbed medium, as well as to determine the nature of state's periodicity and whether the system is dissipative. The system will be considered unstable in the conventional sense that there is at least one mode of disturbance to the shock parameters that grows during a finite period of time, so the system progressively departs from its initial/stationary state and never reverts to it [14]. To determine the instability locus that separates stable and unstable states, a set of parameters $\Sigma\{x_1, x_2, \ldots x_n\}= 0$ that defines the states of neutral (marginal) stability is to be obtained.

II. **THE MODEL OF THE SHOCK WAVE INSTABILITY ON A PERIODICALLY DISTURBED INTERFACE**

Let an initially planar shock wave steadily moves along the *x*-direction with velocity $V_1$ through a uniform gas of temperature $T_1$, from left to right, toward a previously disturbed discontinuous interface separating a gas of higher temperature $T_2$. The stationary interface disturbance will be modeled with the sinusoidal function of known wavelength and amplitude, $x = A\sin(ky)$, $k = 2\pi/\lambda$ (dashed line in Fig. 1.), that was superimposed on an initially planar interface. Gases on both sides of the interface are assumed to be ideal and the pressure is continuous across the interface. The medium in front of the interface will always be considered as uniform and the temperature step across the interface $T_2/T_1 \equiv T_{21}$ will be kept fixed.

In accordance to [5], the onset of the shock perturbation development will coincide with the beginning of the shock interaction with the disturbed medium (i.e. crossing the interface) and its evolution will proceed in the form of front stretchings into the hotter medium (red curve in the Fig. 1).

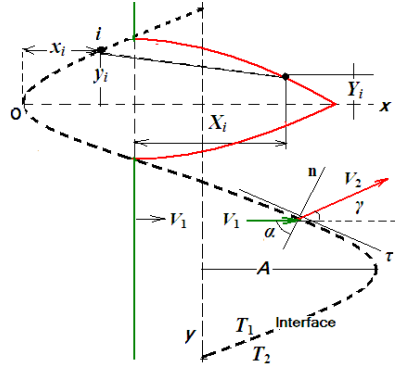

*Fig. 1. The shock perturbation profile (red curve) for one period of the interface disturbance (dashed line), in the vertical plane of symmetry. The initially stable planar shock (green line) is moving from left to right.*

The shock perturbation profile coordinates $(X_i, Y_i)$ at a point of interaction $i$ can be determined using the approach [5] adapted to the system geometry in the Fig.1:

$$X_i = (V_{21} \cos\gamma - 1)(V_1 t - x_i), \quad Y_i = y_i \pm (V_{21} \sin\gamma_i)(V_1 t - x_i) \qquad (1)$$

where $(x_i, y_i)$ are the interface disturbance coordinates $x_i = A[1 \pm \sin(ky_i)]$, $\alpha_i = \tan^{-1}[Ak\cos(ky_i)]$ and $\gamma = \alpha - \tan^{-1}(\sqrt{T_{21}} M_{21n} \tan\alpha)$ are the incident and refraction angles accordingly, $V_{21} = \sqrt{T_{21}(M_{21n})^2 \cos^2\alpha + \sin^2\alpha}$, $V_{21} = V_2/V_1$, $M_{12n} = 1/M_{21n}$, $T_{12} = 1/T_{21}$ are the notations used in the equations, and the rest of the parameters are defined in the Fig. 1. The +/- signs correspond to the lower and upper halves of one period of the interface accordingly and reflect the fact that the shock refraction on each half works in opposite directions, i.e. toward or off their symmetry axes. The Mach number components along the normal to the interface $M_{1n}$ and $M_{2n}$ are defined through the corresponding velocity components [7] and their ratio $M_{21n} = M_{2n}/M_{1n}$ accounts for the losses due to the shock reflections off the interface [13]. In the equations below, the interaction time $t = n_t A/V_1$, $0 \leq n_t \leq 2$ counted from the first moment of the interface crossing and scaled with the characteristic time $\tau = A/V_1$ will appear as the normalized time $\bar{t} = t/\tau = n_t$. Also, because of the symmetry, the derivations will be done only for upper parts of the interface halves (1st and 3rd quarters of the wavelength in the Fig. 1) and the expressions for the rest can be obtained using the $y \to -y$ transformation relative to the symmetry axes. Since the shock refraction effect was shown significantly dependent on the gas density profile behind the interface [11,12], a few the most common types of the distribution will be considered separately.

### III.  A. The uniform case

If the hotter medium is uniform, a portion of the shock front crossing the interface accelerates to velocity $V_2$ in a step-like manner and after this its velocity remains constant. Due to the acceleration, this portion advances faster compared to the reminder that is still in the colder media thus resulting in the continuously increasing front stretching toward the hotter medium [5]. Consequently, the shock loses its stability immediately after it starts to interact with the interface disturbance (point $O$ in Fig.1) and the perturbation to the shock is expected to grow with time during the crossing.

To determine the instability locus, a set of the system parameters defining the onset of the instability growth must be identified. For the two-dimensional shock perturbation $\bar{D}(t,\bar{p})=(\tilde{X},\tilde{Y})$, its differential

$$d\bar{D}(t,\bar{p}) = \frac{\partial \tilde{X}}{\partial t}\partial t + \frac{\partial \tilde{Y}}{\partial t}\partial t + \sum_j \left[\frac{\partial \tilde{X}}{\partial p_j}\partial p_j + \frac{\partial \tilde{Y}}{\partial p_j}\partial p_j\right] \qquad (2)$$

defines the growth rates and the gradients over the system parameters $p_j$ controlling the direction of the instability development. The marginal state of the system discriminating between the stable and unstable states can be obtained from analyzing the first two terms in the equation, the growth rate component $g_x = \partial X_i/\partial t$ determining the rate of the longitudinal stretching, and $g_y = \partial Y_i/\partial t$ that is responsible for the transversal distortion and ultimately the instability decay rate.

In accordance to (1), the uniform parameter distribution results in the $x$-component of the growth rate $g_x = \partial X_i/\partial t$ being time-independent and the requirement for the instability continuous growth during the time of crossing the interface $g_x = (V_{21}\cos\gamma - 1)V_1 > 0$ reduces to

$$V_{21}\cos\gamma > 1 \qquad (3)$$

where $V_1$ and $V_2$ in the ratio are the absolute values of the velocities, and $0 < |\gamma| < \pi/2$. The most significant growth of the shock perturbation determining the overall degree of the instability development will be located behind the upper half of the interface disturbance period due to the interaction times being twice as long compared to the lower one. Thus reaction of the system to the upper half only will be analyzed assuming that the contribution of the lower one will always lag behind and thus will not determine the fate of the instability development.

When evaluating the condition (3) analytically in terms of the system parameters, the expression appears too bulky even with simplifications coming from considering a smooth interface that eliminates the Mach number ratio $M_{21n}$ [5]. It can be noticed though that for axisymmetric problems of similar geometry the most significant changes in the front are known to occur in the vicinity of the symmetry axis [11,12] because of longer interaction times and smaller incidence angle α. Since the maximum of the shock perturbation growth determines the tendency in the instability development, a local solution to the condition can be utilized without much loss of generality. Then, in the small angle approximation $\alpha \to 0$, the trigonometric factor in (3) $\cos\gamma \approx \left(1+\sqrt{T_{12}}\alpha^2\right)/\sqrt{1+\alpha^2(1+T_{12})}$ is always positive and when considering together with the condition $T_{12} \ll 1$, it can be safely approximated with the unit. Another factor in (3) $V_{21} = \sqrt{T_{21}(M_{21n})^2\cos^2\alpha + \sin^2\alpha} > 1$ whenever $\sqrt{T_{21}}M_{21n} > 1$. With $M_{21n} = M_{21}\cos(\alpha-\gamma)/\cos\alpha$ and considering only small α, the last inequality reduces to

$$\sqrt{T_{21}}M_{21} > 1 \qquad (4)$$

Thus the requirement (3) is fulfilled if the condition (4) holds that, in case of a smooth interface type ($M_2 = M_1$), further reduces to $T_{21} > 1$. In (4), the minimum heating intensity $T_{21}$ requirement is dependent on the losses due to shock reflection off the interface accounted through the Mach number ratio and thus will be greater than the unit if the interface is not perfectly smooth [5]. In addition, since the model inherently assumes a non-zero front-to-interface curvature χ that gives rise to a non-zero angle α, it should be added to the condition (4). Thus the full requirement

$$\{T_2/T_1 \geq (M_1/M_2)^2, |\chi| \geq 0\} \tag{5}$$

finally defines the set of system parameters determining the marginal state. Here, the equal sign is applicable to the instability locus and the > sign determines the direction in the parameter change to trigger the instability.

Similarly, the transverse component of the growth rate can also be found time-independent and the instability continuous growth in this direction requires $g_y = \partial Y_i/\partial t = V_1(V_{21}\sin\gamma) > 0$ that, in accordance with the above conclusions, is also satisfied.

In investigating the factors influencing the mode of the instability evolution, the growth rates dependence on the interface disturbance parameters can be examined. In the same small angle approximation, i.e. within the range of small distances $\Delta y$ from the symmetry axis, and for upper half of the interface $ky = (\pi/2) \pm k\Delta y$ so the rates

$$g_x \approx V_1\left[\left(\frac{T_{21} + 3(Ak^2\Delta y)^2}{1 + (2+T_{12}^2)(Ak^2\Delta y)^2}\right)^{1/2} - 1\right], \quad g_y \approx V_1\left(\frac{Ak^2\Delta y T_{21}(1-\sqrt{T_{12}})}{1 + Ak^2\Delta y(1-T_{12})}\right)^{1/2} \tag{6}$$

can be found dependent on the heating strength across the interface disturbance and the geometrical factor $Ak^2$. The physical meaning of the factor becomes clear if one tries relating it to the interface disturbance curvature. For a sinusoidal shape of the disturbance $f(y) = A\sin(ky)$, its curvature conventionally defined as $\chi = |f''|/(1+f'^2)^{3/2}$ can be found

$$\chi = Ak^2\cos(k\Delta y)/(1 + A^2k^2\sin^2(k\Delta y))^{3/2} \tag{7}$$

and approximated with $\chi \approx Ak^2 \propto A/\lambda^2$ in the proximity to the symmetry axis. Thus, with $T_{21}$ fixed, the perturbation growth rates are determined solely by the perturbation curvature $\chi$ rather than by its geometrical parameters $A$ and $k$ separately. There is no critical value of the curvature as its value of any smallness will trigger the instability, though it's growth rate will be shown to be dependent on this parameter.

The similarity law found here points at the phenomenological origin of the shock instability, with the similarity parameter χ as the key factor in the instability triggering, while the heating intensity $T_{21}$ and the initial state shock strength $M_1$ are the parameters controlling the rates and ultimately the degree of the instability development. Consequently, within the used approximations, the interface disturbances of different dimensions but of similar shape (corresponding to the same curvature) should also produce the shock perturbations of similar shapes.

To illustrate the conclusions, the instability development of an initially planar shock wave due to a small waviness of the interface was simulated using the system of equations (1). The initial Mach number was taken as $M_1 = 1.9$, the interface parameters $A = 0.20$ cm, λ= 4.0 cm, $T_{12} = 0.10$, and for stronger effects the interface type was considered smooth. The Fig. 2 demonstrates the shock perturbation advancement into the hotter medium in the vertical plane of the interaction. The curves correspond to times ranging between $n_t = 0.25$ and 2.5 through the 0.25 intervals.

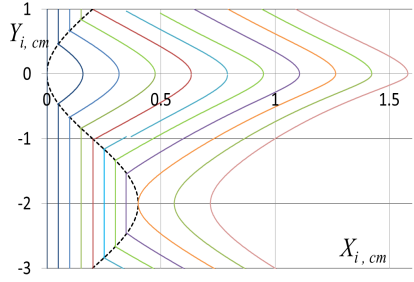

Fig. 2. *The shock front instability development for one period of the interface disturbance, for the interaction times between $n_t$ = 0.25 and 2.5 through the 0.25 intervals. $T_{12}$ = 0.10, $M_1$ = 1.9, $A/\lambda$ = 0.05 (A = 0.20 cm, $\lambda$ = 4.0). The initially planar shock moves from left to right.*

In agreement with the results of analysis in (5), the shock perturbation experiences continuous growth with time thus determining the mode of the perturbation evolution in this case as aperiodical. After the crossing the interface disturbance, the perturbation begins its free motion in the uniform medium with constant speed while retaining the same shape as the conditions for the shock refraction cease. Thus, at the moment of exiting the interface disturbance, the shock wave reaches a new stable state characterized by a new front structure and the gas parameter distribution behind it. The interaction time with the interface (i.e. the time of its full crossing) $t_{int} = 2A/V_1$ determines the time of transition to another state of stability.

However, the possibility of this kind of reorganization of the steady-state flow can be limited in the case of a stronger refraction effect resulting in the front overstretching and decay via degeneration of the shock into a sonic wave [9]. This can occur because, while the shape of the perturbation remains similar in accordance to the law (7), the perturbation's absolute value is still dependent on the disturbance amplitude through the interaction time $t_{int} = 2A/V_1$ that can be long enough to have the shock perturbation overdeveloped and vanished.

Within the simplified conditions imposed on the model, the perturbed shock moving in the uniform medium behind the interface disturbance is supposed to stay in its new stable state indefinitely. But in the broader reality the question about the shock's stability in this post-interaction region can remain open. In accordance to [2,15,16] a small sinusoidal disturbance to an initially planar shock front in a uniform medium free of viscous damping effects can oscillate with an amplitude that decays with time asymptotically, so the shock can regain its planarity. This may leave a possibility for the shock under investigation to eventually return to its initial planar state. In deciding whether it is possible, the smallness [1,2] and the nature of the disturbance [2], as well as whether the medium obeys an ideal-gas equation of state [16] must be considered.

The similarity effect (7) was numerically verified using simulations for the shock perturbation profiles performed for different combinations of the interface disturbance amplitude and wavelength but keeping the small ratio $A/\lambda$ same (corresponding to the same curvature). The results for several sets of parameters $(A/\lambda)$ including (0.1/1.0), (0.2/2.0), (0.3/3.0), (0.4/4.0), (0.5/5.0) returned the curves of identical shapes thus confirming the above conclusions. While observing the similarity in the shape, the dimensions of the perturbation were still dependent on the disturbance amplitude *A* because of the time factor present in the eq. (1).

The effect of curvature on the mode and the degree of the instability development is comparatively illustrated by letting the shock numerically interact with the interface disturbances of two different curvatures, as shown in the Fig.3. For this, the wavelength was kept fixed ($\lambda = 4$ $cm$) and the disturbance amplitude was varied, with $A = 0.1$ $cm$ corresponding to the graph (a) and $A = 0.4$ $cm$ - to the graph (b). The curves on both graphs correspond to the same times $n_t = 0.1$-$1.0$ with $\Delta n_t = 0.1$ intervals and other system parameters were kept the same as for the results in Fig.2.

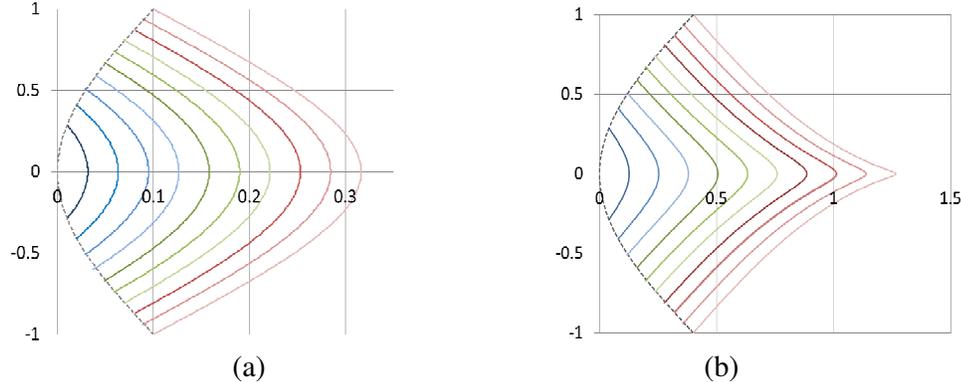

(a)  (b)

*Fig. 3. The effect of curvature on the degree of the shock instability development for the upper half of the interface disturbance, for the case of a lower curvature (a) ($\lambda =4$ cm, $A=0.1$ cm) and a higher curvature (b) ($\lambda =4$, $A=0.4$ cm).*

The initially planar shock is incident from left to right and its planar portions that are still outside of the hotter medium are not shown in the picture. The data is presented only for the upper part of the interface disturbance that is leading in the instability development. The shorter times $n_t$ between 0.1 and 1.0 were chosen to avoid the front perturbation overstretching and collapse that is about to occur at the last moment in the Fig. 3(b). As seen in the figure, the curvature increase brings the qualitative difference to the degree of the instability development that occurs for the same time interval. Stronger refraction effect leading to the increased transversal compression of the perturbation works through the increased growth rates (7) and the longer interaction time $t = n_t (A/V_1)$. The approaching of dissipation can be predicted with the appearance of the inflection points on the curves and is expected to occur faster on a more curved interface disturbance.

As seen, the instability described here develops not in the conventional sense, i.e. when a short-time disturbance to the shock would be applied and then the shock is allowed to evolve with time. In the present case, the disturbance to the shock is applied continuously during the entire time of crossing the interface. In this sense, the instability could be referred to the same type as, for example, Rayleigh-Taylor instability that develops under continuously applied acceleration, though the nature of that instability is different because the shock refraction is not involved there. The front perturbation evolution, though, still occurs in the conventional sense, as the shock structure departs considerably and irreversibly from its initial stable state (planar shock) with the possibility of transition to another stable state or decay.

### II.B. The density distribution is non-uniform

In the case of a non-uniform density distribution, different portions of the front perturbation propagate in the medium with different local temperatures and hence the local velocities of these portions are also different. Thus the instability will continue to evolve even after crossing the interface, for as long as the distribution is supported or until the instabilities dissipate. The way the instabilities develop during this time is determined by the type of the density distribution. But in any case, because of the non-uniformity, the time dependence of the growth rates can result in qualitatively different outcomes for the instability development.

### B1. The density is decreasing exponentially

The exponential type of density distribution that can be established due to the exothermal expansion is common for such important problems as strong explosion in the earth's atmosphere [17], large-area plasma sources created with internal low-inductance antenna units [18], detonation [7], or the ultra-intense laser-induced breakdown in a gas [19]. The density will be considered decreasing in the longitudinal direction only, starting from a finite value $\rho_{00}$ at the interface in accordance with the law $\rho_2(x) = \rho_{00} \exp(-x/z_0)$, and its change in the transverse direction can be neglected if the interface waviness amplitude is small compared to the distribution characteristic length $z_0$, $z_0 \gg A$. The media on both sides of the interface is assumed to be ideal gases and the gas pressure across it continuous, so the relation $\rho_{12} = T_{21}$ is held. During crossing the interface, the shock perturbation profile can be determined with the system of equations similar to that derived in [12]

$$X_i = \sigma\left(t - t_{0i} + t_\lambda\right)^{2/5} + \varepsilon\left(t - t_{0i} + t_\lambda\right)^{4/5} - x_0, \quad Y_i = y_i \pm \left[\sigma\left(t - t_{0i} + t_\gamma\right)^{2/5} - \sigma\, t_\gamma^{2/5}\right] \quad (8)$$

where $t_{0i} = x_i/V_1$ is the delay time (to the interface), the parameters

$$x_0 = \sigma t_\lambda^{2/5} + \varepsilon t_\lambda^{4/5} \quad \text{and} \quad t_\gamma = \left(\frac{5V_1\sqrt{\rho_{12}}\cos\alpha\cdot\sin\gamma}{2\sigma\cos(\alpha-\gamma)}\right) \quad (9)$$

$\sigma = \xi(E/\rho_{00})^{1/5}$, $\varepsilon = (K/z_0)\sigma^2$ are defined through the effective explosion energy $E$, $\xi = 1.075$ and $K = 0.185$ are the numerical parameters from [20], $t_\lambda$ is found from the solution of the following equation

$$\frac{5}{2}\frac{V_1 \cos\alpha}{\sigma\cos(\alpha-\gamma)}\sqrt{\rho_{12}} = t_\lambda^{-3/5} + \frac{2\varepsilon}{\sigma}t_\lambda^{-1/5}, \quad (10)$$

and the rest is defined in Fig.1.

The instability locus is determined by evaluating the growth rate components

$$g_x = \frac{2}{5}\left[\sigma\left(t - t_{0i} + t_\lambda\right)^{-3/5} + 2\varepsilon\left(t - t_{0i} + t_\lambda\right)^{-1/5}\right], \quad g_y = \frac{2}{5}\sigma\left(t - t_{0i} + t_\gamma\right)^{-3/5} \quad (11)$$

that are positive as the condition of crossing the interface assumes the time $t$ be larger than the delay time $t_{0i}$, and $t_\lambda$ and $t_\gamma$ are the positive parameters if $|\alpha| < \pi/2$, $|\gamma| < \pi/2$. Since the requirement for the locus is determined by the conditions on the interface only, the same set of critical parameters (5) is applied here to trigger the instability.

The results of numerical simulation for the shock perturbation evolution ($X_i$, $Y_i$) based on the system (8) are presented in the Fig.4, for the times between $n = 0.3$ to 3.1 through the equal intervals $\Delta n = 0.3$. The interface parameters were taken as $A = 0.3$ cm and $\lambda = 4$ cm, $z_0 = 5.0A$, $\alpha = 7$ and $\beta = 402.79$ correspond to the specific explosion energy $E/c_{00} = 11.707 \cdot 10^3$ J m$^3$/kg [20], the type of the interface is smooth, and the rest is the same as in the uniform case. The main difference from the uniform case here is that, while the perturbation to the shock steadily grows during the time of crossing and after this, both components of its growth rate, in accordance to (11), are decreasing functions of time for all the interaction points. Graphs in the Fig. 5 demonstrate the saturation tendency for the rates that occur at times corresponding to finishing crossing the interface disturbance. The non-zero saturation levels of the rates varying very slowly for the most locations on the disturbance surface still ensure the continuous shock perturbation growth with time. The behavior of the curves is determined by the parameters $t_\lambda$ and $t_\gamma$. With the shock and gas parameters fixed, they are only geometry dependent and thus, as seen in the graphs, the behavior is very location sensitive.

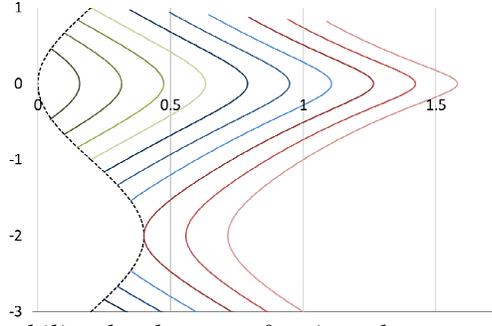

*Fig. 4. The shock instability development for times between $n_t = 0.3$ and $3.1$ through the intervals $\Delta n_t = 0.3$. The gas density behind the interface is exponentially decreasing. All dimensions are in cm, $A = 0.3$ cm, $\lambda = 4$ cm.*

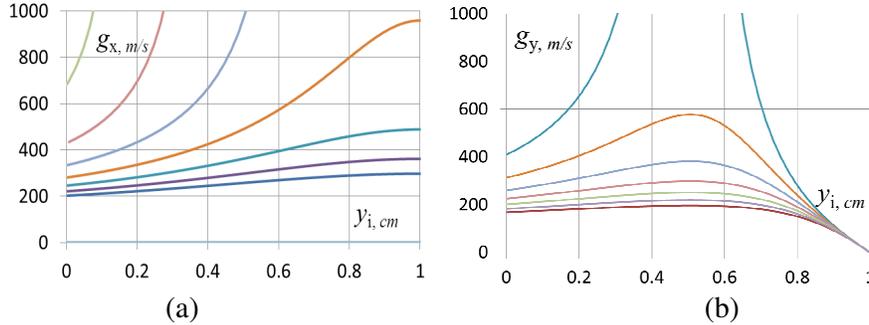

*Fig. 5. The $g_x$ (a) and $g_y$ (b) components of the growth rate versus $y_i$, for the same interaction times as in Fig.4. $A = 0.3$ cm, $\lambda = 4$ cm. The time sequence for curves is from upper to lower.*

The interface parameters $A$ and $k$ affect the shock perturbation structure through variations in the incidence angle and the delay time present in the dependencies for the parameters $t_{0i}$, $t_\lambda$, and $t_\gamma$. In the small angle approximation, $T_{12} \ll 1$, and for upper half of the interface, the parameters can be approximated using the same discrete analysis as in paragraph IIa

$$t_{0i} = (A/V_1)[1 - \cos(k\Delta y)] \approx (Ak^2)\Delta y^2 / 2V_1 \tag{12}$$

$$t_\gamma \approx \left(5V_1 \sqrt{T_{12}} / 2\sigma\right)\left(1 - \left((Ak^2)\Delta y\right)^2\right)\sqrt{\frac{(Ak^2)\Delta y(1 - T_{12})}{1 + (Ak^2)\Delta y(1 - T_{12})}}, \quad T_{12} \ll 1 \tag{13}$$

Since the ratio $2\varepsilon/\sigma$ is a constant factor, the $t_\lambda$ dependence on the interface parameters can be determined from analyzing the left hand side term of the equation (10)

$$\frac{5}{2}\frac{V_1 \cos\alpha_i}{\sigma \cos(\alpha-\gamma)}\sqrt{T_{21}} \approx \frac{5V_1}{2\sigma}\left(1-\frac{(Ak^2)^2 \Delta y^2}{2}\right)\sqrt{T_{21}+(Ak^2)^2 \Delta y^2} \quad (14)$$

As seen in (12-14), the common dependence of the parameters $t_{0i}$, $t_\lambda$, and $t_\gamma$ on the same factor $\chi \approx Ak^2$ establishes the similarity law with respect to the curvature in this non-uniform case too. To visualize this conclusion, numerical simulations of the instability development were run for a set of different parameters $A$ and $\lambda$ but their ratio kept the same. In the 10-fold range of the parameter values ($A/\lambda = 0.1/2.0 \div 1.0/20.0$), the entire shock perturbation profiles were of identical shapes, similar to that in Fig. 4. The results thus suggest that even though the conclusions (12-14) were derived in the vicinity of the symmetry axis, they seem to be valid on the periphery as well.

### B2. The density is decreasing with the Power Law

When the gas density is decreasing to zero at a distance $a$ in accordance with the power law $\rho \propto x^N$, a moving through such a medium strong plane shock can accelerate very quickly accumulating virtually infinite energy (cumulation effect) [21,18,22]. For a small $A/a$ ratio, the density can be assumed as changing off the interface, in the $x$- direction only (along the symmetry axis). For the present geometry, the shock perturbation profiles can be found utilizing solutions from [11]

$$X_i = -(G/b)(t_{0i}-t)^b + a_i, \quad Y_i = y_i - (V_{2y}/V_1)(V_1 t - x_i) \quad (15)$$

where $a_i = a - x_i$ is the local horizontal distance to the zero density plane from the interface, $V_{2y} = V_2 \sin\gamma$, and $t_i = x_i/V_1$ is the local delay time. The local travel time to the zero density plane

$$t_{0i} = \frac{b(a - A(1 \mp \sin(ky)))}{V_1 \cos\gamma\sqrt{T_{21}M_{21n}^2 \cos^2\alpha + \sin^2\alpha}} + \frac{x_i}{V_1}, \quad G = \frac{V_1 \cos\gamma\sqrt{T_{21}M_{21n}^2 \cos^2\alpha + \sin^2\alpha}}{[t_{0i}-(x_i/V_1)]^{b-1}}, \quad (16)$$

and the constant $b = 0.59$ [22].

The instability locus is determined from analyzing the inequality $G(t_{0i}-t)^{b-1} \geq 0$ that transforms into

$$\frac{V_1 \cos\gamma_i \sqrt{T_{21}M_{21n}^2 \cos^2\alpha + \sin^2\alpha}}{[t_{0i}-(x_i/V_1)]^{b-1}}(t_{0i}-t)^{b-1} \geq 0 \quad (17)$$

It is satisfied under the following set of simultaneous conditions: **(a)** $t_{0i} \geq t$ that is always the case as at $t = t_{0i}$ the perturbed shock approaches the zero density location [22]. **(b)** $t_{0i} \geq (x_i/V_1)$ is satisfied if the travel time to the zero density plane is always larger than the delay time. Associated with this inequality $a_i = a - A(1 - \sin(ky)) \geq 0$ essentially requires the interface disturbance amplitude to be within this particular density distribution limits. **(c)** The requirement $\cos\gamma \geq 0$ is true for $|\gamma| \leq \pi/2$, i.e. assuming that the disturbed front does not reverse the direction of its motion.

Since the constant $b < 1$, the component of the growth rate $g_x$ is a very quickly increasing function of time. It also features a high acceleration rate of the increase thus ensuring a quick and non-linear instability growth. Thus the shock perturbation growth in this case is also unconditional and the shock instability can be classified as absolute.

In the vicinity of the symmetry axis, the growth rate dependence on the interface disturbance parameters for upper half of the interface can be approximated as

$$g_x(A,k,t) = \left(\frac{V_1}{ab}\right)^{b-1} \left(\frac{T_{12} + (Ak^2\Delta y)^2}{1 + (Ak^2\Delta y)^2}\right)^{\frac{b}{2}} \left[\left(\frac{ba}{V_1}\right)\sqrt{\frac{1 + (Ak^2\Delta y)^2(1 + 2T_{12})}{T_{12} + (Ak^2\Delta y)^2}} - t\right]^{b-1} \quad (18)$$

and, as in both previous cases of the density distribution, displays the similarity feature with respect to the interface curvature $\chi \approx Ak^2$.

Presented in Figs. 6 a,b, the simulation results are obtained for two strengths of the density distribution that was varied with the length $a$ in the range $a = (10 \div 15)A$, $a \gg A$. The interface disturbance parameters $A = 0.2$ cm and $\lambda = 4$ cm for both graphs, and the rest is the same as for Figs. 2 and 4. Note the different interaction times in the graphs (a) and (b) of Fig.6 used to keep the instability development to approximately the same degree under the conditions of different density distribution strengths. As seen in the graphs, the substantially quicker growth of the shock perturbations, their advancement deeper into the hotter medium, and, as a consequence, reaching the dissipation phase faster, are the features common for this particular density distribution.

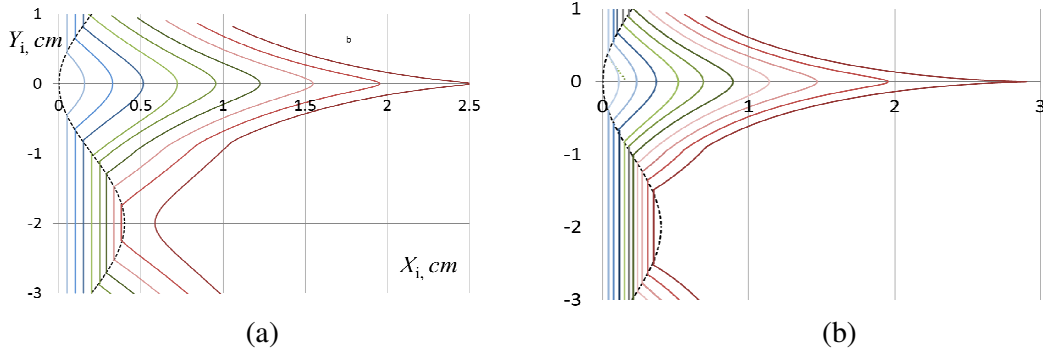

(a)                  (b)

Fig. 6. *The shock instability development for the power law density distribution case. (a) The density gradient is of a moderate strength ($a = 15A$). The curves correspond to times starting at $n_t = 0.24$, through the intervals $\Delta n_t = 0.24$. (b) The density gradient is stronger ($a = 10A$). The times are starting at $n_t = 0.17$, through the intervals $\Delta n_t = 0.17$. The parameters $A=0.2$ cm and $\lambda=4$ cm are the same for both graphs.*

While for a moderate density gradient modeled with a longer distance $a = 15A$ (Fig. 6 a) the shock perturbation profiles still resemble those obtained for the uniform and exponential density distributions, in the stronger gradient case ($a = 10A$, Fig. 6 b), the outcome changes qualitatively. In its upper portion, the shock perturbation develops at a so high rate that it reaches overstretching well before its lower portion, still propagating with the lower velocity $V_1$, even finishes crossing the interface. Due to the outpacing of the upper portion over the lower one, there is a possibility for one part to dissipate while the other is still interacting with the interface. This results in the broken front configuration, with "holes" corresponding to the protruding parts of the interface disturbance.

The critical relationship between the parameters, when the "hole' starts to appear, can be estimated considering the local shock front inclination angle $\varphi$ approaching its zero value (in the upper portion) at the time when the lower one just finishes crossing the interface disturbance

$$\tan\varphi = V_{21} V_1 Cos\gamma (t_{0i} - 2A/V_1)^{1-b} / G_i \quad (19)$$

In the current geometry, the dissipation occurs first in the vicinity of the symmetry axis (Fig. 6), so in the small angle approximation

$$\varphi \approx (Ak^2\Delta y/a^{1-b})(\sqrt{T_{21}}M_{21}/b)^{1-b}(\sqrt{T_{21}}M_{21} - 1)[ba/(\sqrt{T_{21}}M_{21}) - 2A]^{1-b} \quad (20)$$

For a non-zero $\Delta y$ and accounting for the condition (5), $\varphi$ reaches the zero when the relative strength of the density gradient exceeds its critical value

$$A/a \geq b/\left(2\sqrt{T_{21}}M_{21}\right) \tag{21}$$

For $T_{21} = 10$, and with $M_{21} = 0.73$ typical in the vicinity of the axis, the estimation gives the ratio $a/A \approx 7.78$. This agrees well with the value around 8.0 at which the phenomenon starts to appear when the equations (15-16) were run numerically.

The relatively small interface disturbance amplitudes $A$ used in the simulations, being typically an order of magnitude smaller than the wavelength $\lambda$ (0.2/4), suggest that, in the presence of a strong density gradient, even a very small interface waviness can result in perturbations of considerable amplitude growing at a high acceleration rate. Quicker instability development leading to its dissipation stage earlier substantially limits the instability lifetime and, together with high non-linearity of the distribution, leaves no opportunity for the shock to transition to another stable state. Thus, regardless of the strength of the distribution, the instability development proceeds in the form of a secondary flow only.

## IV. INSTABILITY OF THE FLOW BEHIND THE SHOCK

Pressure perturbations caused by the front stretching result in the changes in the flow behind it characterized by a specific parameter re-distribution [20]. The flow state can change, for example, from being uniform to substantially non-uniform [17], or become highly structured and eventually organize into an intense vortex of considerable size [6]. To explore if a small sinusoidal waviness of the interface is capable of inducing the flow instability, the macroscopic model developed earlier in [6] will be applied here. The power law density distribution will be considered here as it was shown supporting the strongest effect, and because of the reasons discussed above only upper half of the interface is of interest. The system of equations (15-16) can be used to compute the vorticity $\omega_i = \omega(X_i, Y_i) = \left(\partial v_x/\partial y - \partial v_y/\partial x\right)_{x=X_i, y=Y_i}$ generated right behind the perturbed front, where

$$v_x = v_n \sin\varphi + V_2 \cos(\phi - \gamma) \cdot \cos\varphi \tag{22}$$
$$v_y = v_n \cos\varphi - V_2 \cos(\phi - \gamma) \cdot \sin\varphi$$

are the components of the flow velocity, $v_n$ is it's normal to the perturbed front component, and $\varphi$ is the local angle between the tangential line to the shock perturbation and the $x$-direction.

The simulation results for the vorticity presented in the Fig. 7 were obtained for the same system parameters and a moderate density distribution strength ($a = 15A$) used in the Fig.6a. The vorticity was scaled with the ratio $V_1/A$ resulting in its dimensionless equivalent $\overline{\omega}_i = \omega(X_i, Y_i)/(V_1/A)$. Slightly shorter interaction times $n = 0.215$ through 2.15 with intervals $\Delta n = 0.215$ were used here to keep the quickly growing vortex intensities within the same graph scale. Due to the symmetry, only one-half of the diagram is shown and the time sequence for the curves is from lower to upper.

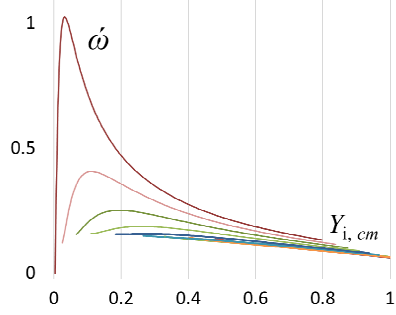

Fig. 7. *Dimensionless vorticity $\bar{\omega}_i$ induced by a small waviness of the interface versus $Y_i$, for the power law density distribution case. A = 0.2 cm, λ = 4 cm, a = 15A. The curves correspond to times between n = 0.215 and 2.15, through intervals Δn=0.215. The time sequence for the curves is from lower to upper.*

Matching the shape of the perturbation profiles in the Fig. 6a with the curve's maximum locations in the Fig.7 suggests that the most intense vorticity developing in the flow corresponds to the areas where the sharpest changes in the front perturbation inclination angle are present, particularly at the locations of the most extreme stretching of the front. The vorticity distribution points at the presence of a clock-wise rotating toroidal vortex (or a roll sheet, depending on the symmetry type), with the curve's maximum corresponding to its center. The maximum's locations and the tendency to shift toward the symmetry axis with time confirm that the next to the axis portion of the front perturbation is the leading part in the instability triggering and its following evolution. Strong dynamics in the vorticity development demonstrate the essentially non-linear character of the flow instability evolution that becomes more remarkable at later times of crossing the interface disturbance. This agrees well with timing in the vorticity development in many experiments where it becomes remarkably visible closer to the end of the interaction. The widths under the curves around the intensity maximums show that the diameter of the vortex ring is on the same scale with the interface characteristic length (λ) and does not change significantly with time.

Thus the results show that the shock perturbation growth in the form of progressive front stretchings into the hotter medium is the only first phase of a complex phenomenon and it is immediately followed by the instability of the flow behind it. The lifetime of the shock-flow instability in this case is limited by the finite distance *a* at which the shock stops [22] or its dissipation, whichever comes first. Due to the direct relationship between perturbations to the shock and to the flow parameter distribution behind it [6], the similarity feature with respect to the curvature found earlier can be extended to the vorticity growth rate as well.

## V. THE INTERFACE STABILITY PROBLEM AND CONCLUSIONS

The origin and the features of the complex shock-flow instability due to a small waviness of the interface were studied in this work. The main finding here is that the shock refraction mediating the interaction between the shock and the interface waviness can be the only mechanism of the complex shock-flow instability. It was shown that the interface disturbance of any smallness triggers the shock instability in the form of stretchings linearly or non-linearly growing with time into the medium of lower density. Pressure perturbations caused by the stretchings result in the loss of stability of the flow behind it that eventually organizes into an intense clock-wise rotating vortex structure. Depending on the density distribution, a transition to another stable state characterized by new geometry and flow parameter distribution can occur, or the shock-flow perturbations can continue to evolve until it dissipates. In accordance with the model, the time of the transition between two states of stability in the uniform density case is equal to the total shock-interface disturbance interaction time. In the non-uniform density case, the transition in the form of front stretching exhibits the pattern of motion that prevails the principle of exchange of stabilities so the instability sets in as a secondary flow. The pattern of the instability development during all the time of crossing the interface disturbance can be classified as absolute and aperiodical (in time, for one interface disturbance period). Considering a periodical interface of an extended length, the final picture of the flow instability will appear as a periodical series of the shock front stretchings followed with vortexes behind each protruding part of the interface disturbance.

The marginal state condition (5) $\{T_2/T_1 > (M_1/M_2)^2, |\chi| > 0\}$ is the only requirement to trigger the instability. It is not density distribution dependent and is determined only by the conditions on the interface disturbance. The minimum heating intensity requirement accounts for losses due to shock reflections off the interface through the Mach number ratio and thus is larger than the unit. The similarity law found valid in the vicinity of the symmetry axis determines the interface disturbance curvature as the single complex parameter governing the instability growth rates rather than the geometrical parameters of the interface disturbance separately. While the gas density distribution behind the interface does not play a role in the instability triggering, it may still critically contribute to its growth and decay rates.

The specific wave nature of the instability dissipation, when the overstretched shock perturbation decays via the degeneration into an acoustic wave, allows the shock instability to decay even though the viscous damping mechanisms are not present in the flow (except the shock width layer). The shock wave energy, in this case, converts into rotational energy of the flow thus continuously supporting the developing vortex structure until the perturbed shock vanishes.

Considering the complex instability in a broader context, the flow instability can be found as not the last step in the interaction. Since the shock and the flow instabilities take place upstream from the interface, the perturbations to the flow parameters propagating downstream, toward the interface, will disturb it. The overall pressure drop behind the perturbed (refracted) shock as a result of the flow parameter re-distribution [9] continuously mounting with time will be responsible for the sucking effect resulting in the large-scale interface perturbation moving it closer to the shock. The positive and essentially non-linear dynamics in the pressure perturbation evolution [9] will support amplification of this global perturbation to the interface. The vorticity generation is another, typically a local consequence of the flow parameter re-distribution [6] that superposes vortical smaller-scale perturbations to the interface and thus determines the pattern in the interface instability structure. With increasing distortion of the interface, Kelvin-Helmholtz (KH) shearing instability may start to contribute resulting in the characteristic mushroom shapes of the interface perturbations [23]

In some sense the extended, now three-stage shock-flow-interface instability described here can be referred to the class of Richtmyer-Meshkov (RM) instability problems, though with a few distinct specifics. In the case of conventional RM instability, any perturbation initially present on the interface is known to amplify, and the baroclinic vorticity generation is considered as the basic mechanism of the amplification [23]. On one hand, it agrees with the model presented here as a small waviness of the interface represents that initial disturbance triggering the shock instability followed by the flow and interface instabilities. On another hand, in accordance with the same model, the initial perturbations to the interface may not even be necessary as the chain of the shock-flow-interface instabilities can also be triggered on an undisturbed (planar) interface as long as the incident shock front is curved. In this case, all that is needed for the instability triggering is the relative non-zero curvature between the incident shock and the interface disturbance [12] and the minimum heating intensity, in accordance with the condition (5). Thus, contrarily to other instability types, the interface instability described here is not about amplification of the initial interface disturbance but rather the general consequence of the shock refraction.

Another distinct feature following from the model is that the vorticity generation, whether via the baroclinity or the refraction effect described here, may also not be necessary for the interface instability to develop. The pressure perturbations resulting in the global pressure drop behind the refracted shock alone will still perturb the interface and the positive dynamics common for the flow instability development ensure the interface instability growth. The initial instability pattern will be of a larger scale in this case and the KH-instability turning on at later stages would finally determine the smaller characteristic structure of the instability typically observed in experiments. The question whether the flow parameter perturbations will evolve in the form of significantly developed vorticity or not can be determined, for example, by the density distribution [6]. In some cases [6] the flow parameter re-distribution behind the shock results in the vorticity of either micro-size and/or micro-intensity whose contribution can be neglected. In this sense, the pressure and the vortical perturbations can be considered competing as the locally developing secondary flows in the form of vorticity result in an effective mixing that helps to quickly equalize pressure in different parts of the volume and thus quench (at least partially) the large-scale sucking effect on the interface.

The phenomenon described here agrees well with numerous experimental observations, with the timing in the interactions correctly corresponding to its later stages, when the refracted shock stretchings, vorticity, and the interface distortion/collapse develop enough to become observable. The instabilities structure, size, and rotation direction correlating well with the model predictions suggest that it can be used in interpretation of the experimental results via identification of its characteristic parameters. For example, the stream-wise small-scaled vorticity generated behind the secondary shock observed during the interaction between a bow shock and a spherical low-density pulse-heated bubble [3], from the perspective of this model, could be also explained by the single refraction mechanism. A small initial waviness on the bubble boundary could trigger the shock instability resulting in a number of small-scale stretchings on the refracted front eventually giving rise to a number of same-scale vortices behind them.

The results of this research can be also found useful in the problems of the shock–flame interactions, front separation regions control experiments, in combustion, and in the electric discharge, RF- or laser-induced energy deposition experiments. The research topic of fluid/plasma dynamics instability and turbulence in impulsively loaded flows can also be of considerable interest in astrophysics plasmas and fusion research.